# Unveiling the Complex Jet Dynamics in the Blazar 2021+317 through Multiepoch Very Long Baseline Interferometry Observations

Haitian Shang[1,2,3], Wei Zhao[1], Xiaoyu Hong[1,2,3,4], Leonid I. Gurvits[1,5,6], Ailing Zeng[1,2], Tao An[1], and Xiaopeng Cheng[7]

[1] Shanghai Astronomical Observatory, Chinese Academy of Sciences, 80 Nandan Road, Shanghai 200030, People's Republic of China; htshang@shao.ac.cn, xhong@shao.ac.cn
[2] School of Physical Science and Technology, Shanghai Tech University, 100 Haike Road, Pudong, Shanghai 201210, People's Republic of China
[3] School of Astronomy and Space Science, University of Chinese Academy of Sciences, Beijing 100049, People's Republic of China
[4] College of Physical Science and Technology, Xiamen University, Xiamen 361005, People's Republic of China
[5] Joint Institute for VLBI ERIC, Oude Hoogeveensedijk 4, Dwingeloo, 7991 PD, The Netherlands
[6] Faculty of Aerospace Engineering, Delft University of Technology, Kluyverweg 1, Delft, 2629 HS, The Netherlands
[7] Korea Astronomy and Space Science Institute, Daejeon 34055, Republic of Korea

Received 2025 March 23; revised 2025 April 30; accepted 2025 May 17; published 2025 July 7

## Abstract

We present an investigation of the compact structure of the active galactic nucleus 2021+317 based on multiepoch very long baseline interferometry (VLBI) observations at 15, 22, and 43 GHz in the period from 2013 through 2024. The VLBI images show a core–jet structure extended to the south, with two stationary components in the northern region, one of which is likely to be the core of the source. We also detected two new moving jet components (S4 and S5) in the observations of 2021. Based on these observational findings, we analyzed two distinctive jet models involving one or another stationary component mentioned above as the jet core. One model assumes a moderate bulk motion velocity, a wider viewing angle, and a lower Doppler factor, with the magnetic field energy density significantly dominating over the nonthermal particle energy density. The other model involves a higher bulk motion velocity, a narrower viewing angle, and a higher Doppler factor, with an even greater dominance of magnetic field energy in the core. The position angle of the jet ridgeline rotates counterclockwise over the observed period. The apparent kinematics of the jet components is more consistent with a model of the precessing jet, which has recently completed the first half of the precession cycle. Our results provide constraints on the dynamic evolution of the jet and its interaction with the surrounding medium.

*Unified Astronomy Thesaurus concepts:* Active galactic nuclei (16); Very long baseline interferometry (1769); BL Lacertae objects (158)

## 1. Introduction

Blazars constitute the most extreme subclass of active galactic nuclei (AGNs). Their relativistic jets are directed at a very small angle from the observer line of sight and have a strong relativistic beaming effect (C. M. Urry & P. Padovani 1995; A. B. Pushkarev et al. 2017). Blazars exhibit significant variability across all bands of the electromagnetic spectrum, with timescales ranging from minutes (J. Albert et al. 2007) to years (M.-H. Ulrich et al. 1997). These variations are attributed to various intrinsic to the source phenomena, including perspective changes of the emission region in a twisted or wobbling jet (C. M. Raiteri et al. 2017), shock waves propagating through the jet (A. P. Marscher & W. K. Gear 1985), magnetic reconnection events in the jet (D. Giannios et al. 2009), turbulence in the jet plasma (A. P. Marscher et al. 2008), and changes in the accretion rate of the central supermassive black hole (C. S. Reynolds & M. C. Begelman 1997), as well as propagation effects in the media between the source and observer (B. J. Rickett 1990).

BL Lacertae (BL Lac) objects represent one of the most extreme subclasses of AGNs. While they share many characteristics with other types of AGNs, they exhibit unique properties in terms of gamma-ray emission. In contrast to flat-spectrum radio quasars (FSRQs), BL Lac objects show distinct gamma-ray emission behaviors and can be further categorized based on their synchrotron peak frequencies (C. M. Urry & P. Padovani 1995). Although FSRQs have been extensively investigated, BL Lac objects remain more challenging to study due to their lack of prominent moving features and extended radio structures. Additionally, the determination of redshifts has proven difficult because of their characteristically featureless optical spectra (H. Netzer 2013). As a result of these challenges, the inner jet structures and dynamical properties of BL Lac objects are still not well understood.

The jet wobbling phenomenon, first described by S. L. Mufson et al. (1990) as a deviation or variation of the jet direction relative to the observer line of sight, is relatively common (A. Caproni et al. 2013, 2017; R. Lico et al. 2020). Erratic wobbling (e.g., CTA 102; C. M. Raiteri et al. 2017) is usually attributed to turbulence in the accretion flow or instabilities at the base of the jet. Periodic wobbling is often regarded as evidence for the existence of a binary black hole system (Z. R. Weaver et al. 2022), which induces regular precession of the jet nozzle leading to the wobbling (e.g., 4C+12.50; M. L. Lister et al. 2003). In addition, the disk-jet precession may also be caused by the Bardeen–Peterson effect (S. Liu & F. Melia 2002; A. Caproni et al. 2004) or the Lense–Thirring effect (or frame-dragging effect; M. Liska et al. 2018), resulting in periodic changes in the jet direction that manifest themselves as jet wobbling.

Very long baseline interferometry (VLBI) enables imaging of distant radio sources with milliarcsecond resolution, revealing submilliarcsecond structures that correspond to parsec-scale dimensions throughout the redshift space. Recent millimeter VLBI studies have shown that submilliarcsecond-scale AGN jets exhibit unique properties distinct from their milliarcsecond-scale counterparts, highlighting the complexity and instability of the innermost jet regions (M. L. Lister et al. 2003; M. J. Valtonen et al. 2012; C. M. Fromm et al. 2013b; S. Koyama et al. 2016; J. A. Hodgson et al. 2017; J. Oh et al. 2022; A. Fuentes et al. 2023).

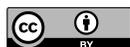

The BL Lac object 2021+317 ($z = 0.356$; M. Ackermann et al. 2015), R.A., decl. (J2000) $20^h23^m19^s.017$, $+31°53'2".306$, is one of the targets in our millimeter VLBI survey, which includes 134 radio-loud AGNs (X. P. Cheng et al. 2018, 2020). X. P. Cheng et al. (2018, 2020) conducted 43 GHz Very Long Baseline Array (VLBA) imaging observations of these sources from 2014 to 2016 (with 10 of the sources also observed at 86 GHz), aiming to perform a statistical study focused on their compactness, core brightness temperature, and correlation between gamma-ray and radio emissions. They found a new off-axis component appearing to the northwest of the source. If this component is the core, then the jet has undergone sudden bending within 1 mas. P. Kharb et al. (2010) found that it has a prominent core–halo structure on kiloparsec scales. The source shows southward-extended jets of 50 mas and 6 mas at 1.4 GHz and 15 GHz, respectively (M. L. Lister et al. 2009b).

Based on the 15 GHz (M. L. Lister et al. 2018) and 43 GHz (X. P. Cheng et al. 2020) VLBI images, we found that the position angle of the jet in this source has undergone significant changes. This change may reflect the dynamical complexity of the core region, providing key insights into the formation and evolution of AGN jets. Additionally, the core is elongated in the southeast–northwest direction, with an elongation length of approximately 0.5 mas, and features an off-axis jet knot (X. P. Cheng et al. 2020). The MOJAVE project conducted 15 GHz observations of this source between 1995 and 2013, finding that the core region (starting from 2010 September 17) could be modeled with two jet components (M. L. Lister et al. 2019), further highlighting the structural complexity of the core. These characteristics make BL Lac object 2021+317 a promising target for studying the physical processes in AGN core regions and the mechanisms driving jet evolution. We plan to further investigate the off-axis knot and the changes in the position angle of the jet in this source.

In this work, we further study the core–jet structure of 2021+317 with multiepoch VLBA observations at 15 GHz, 22 GHz, and 43 GHz. In Section 2, we describe the observations, the archival data, and the data reduction process. In Section 3, we present the resulting VLBI morphological properties of the source, and in Section 4, we give a discussion of our findings. Section 5 summarizes the results and presents our conclusions.

Throughout the paper, we use a $\Lambda$CDM cosmology with $H_0 = 71$ km s$^{-1}$ Mpc$^{-1}$, $\Omega_m = 0.27$, and $\Omega_\Lambda = 0.73$ (E. Komatsu et al. 2011). For 2021+317 with a redshift of 0.356, an angular size of 1 mas corresponds to a projected linear length of 4.96 pc, and a proper motion of 1 mas yr$^{-1}$ corresponds to 24.1$c$. The spectral index $\alpha$ is defined as $S_\nu \propto \nu^\alpha$.

## 2. Observations and Data Analysis

This work is based on observing data obtained in our own observations and available in the VLBA data archive.[8,9] Radio telescopes involved in this study are listed in Table 1. The following subsections describe our own observations and data obtained in the archive.

---
[8] https://data.nrao.edu/portal/#/
[9] https://astrogeo.org/vlbi_images/

**Table 1**
List of Radio Telescopes Involved in Observations Discussed in the Current Work

| Array | Station | Code | Diameter (m) |
|---|---|---|---|
| VLBA | Brewster | BR | 25 |
| | Fort Davis | FD | 25 |
| | Hancock | HN | 25 |
| | Kitt Peak | KP | 25 |
| | Los Alamos | LA | 25 |
| | Maunakea | MK | 25 |
| | North Liberty | NL | 25 |
| | Owens Valley | OV | 25 |
| | Pie Town | PT | 25 |
| | Saint Croix | SC | 25 |
| EAVN | VERA-Mizusawa | MIZ | 20 |
| | VERA-Iriki | IRK | 20 |
| | VERA-Ogasawara | OGA | 20 |
| | KVN-Yonsei | KYS | 21 |
| | KVN-Ulsan | KUS | 21 |

### 2.1. VLBA Observations at 43 GHz

In this work, we present two of our VLBA observations of the source 2021+317 at 43 GHz. The first one took place in 2016 January as part of the survey program described in X. P. Cheng et al. (2018). This observation revealed a new jet component on the northwestern side of the brightest component of 2021+317. In the current work, we reprocessed the data using the same data handling routine as used for the following new observation. The latter was conducted in 2021 April (code: BZ081; PI: Wei Zhao). The parameters of both observations are listed in Table 2. The data rate was 2 Gbps with 2 bit sampling. Both left-hand circular polarization and right-hand circular polarization were recorded, covering a total bandwidth of 256 MHz. The data were divided into four subbands (IF), each with a bandwidth of 64 MHz. We use BL Lac and 3C 454.3 as calibrators, and for every two scans of the target source, we observe the calibrators once.

### 2.2. EAVN Observations at 22 and 43 GHz

In 2023 March, we observed 2021+317 with EAVN at 43 GHz (8 hr) and 22 GHz (4 hr) to bridge the source morphology between millimeter and centimeter wavelengths. The observations were conducted in dual-polarization mode with five antennas: MIZ, IRK, OGA, KYS, and KUS. The maximum baseline length reached 1800 km (OGA-KYS), achieving theoretical angular resolutions of 1.2 mas and 0.6 mas at 22 GHz and 43 GHz, respectively. The theoretical imaging sensitivities were 0.2 and 2.1 mJy beam$^{-1}$ at 22 GHz and 43 GHz, respectively. Its $uv$-coverage is shown in Figure 1. We use BL Lac as the fringe finder, bandpass (BP) calibrator, and right-Left circular polarization (RL) delay calibrator. 3C 454.3 is also used as a fringe finder. In this paper, we use this data set to determine the spectral index of components E and W. The detailed information of these data is listed in Table 2.

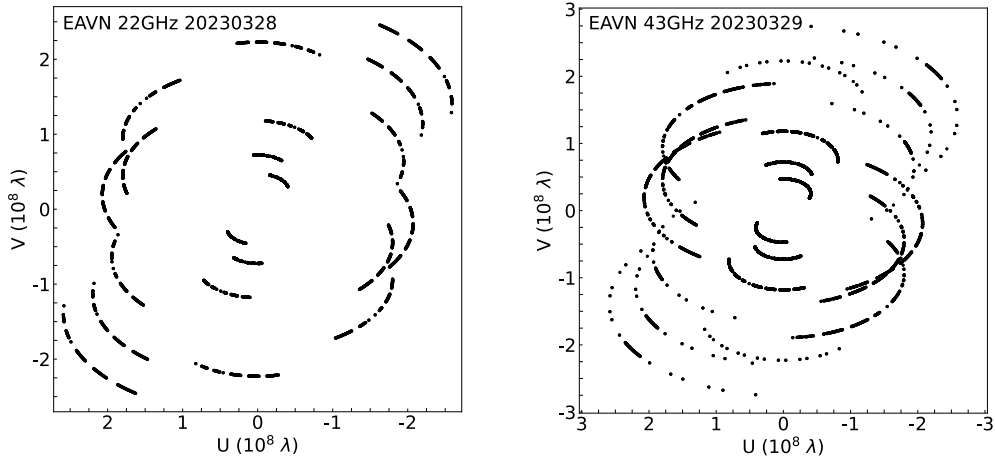

**Figure 1.** *uv*-coverage of EAVN observations at 22 and 43 GHz.

Table 2
Observing Parameters of All Observations (Re)analyzed in the Current Work

| P.C. (1) | Date (2) | Freq. (GHz) (3) | Array (4) | Time (s) (5) | Data Rate (Mbit s$^{-1}$) (6) | Principal Investigator (7) |
|---|---|---|---|---|---|---|
| BL286AA[a] | 2021-08-07 | 15 | VLBA | 1326 | 4096 | |
| BL286AH[a] | 2022-02-24 | 15 | VLBA, –KP | 1348 | 4096 | |
| BL286AO[a] | 2022-09-29 | 15 | VLBA, –KP, –PT | 1336 | 4096 | Matthew Lister |
| BL286AU[a] | 2023-05-27 | 15 | VLBA | 1254 | 4096 | |
| BL286BB[a] | 2023-12-15 | 15 | VLBA | 1394 | 4096 | |
| BL286BG[b] | 2024-06-02 | 15 | VLBA | 1362 | 4096 | |
| BR145YB[c] | 2013-11-30 | 22 | VLBA, –FD, –PT | 145 | 512 | Andreas Brunthaler |
| BM421[c] | 2015-07-07 | 22 | VLBA | 302 | 2048 | James Miller-Jones |
| BR210CD[c] | 2016-02-25 | 24 | VLBA, –SC | 128 | 512 | Mark Reid |
| UD001C[c] | 2017-02-23 | 24 | VLBA | 356 | 2048 | VLBA Operations |
| UD001W[c] | 2018-06-30 | 24 | VLBA, –NL | 177 | 2048 | |
| BZ076A[c] | 2019-03-17 | 22 | VLBA | 56 | 512 | Ingyin Zaw |
| a2301[d] | 2023-03-28 | 22 | EAVN | 14,400 | 1024 | Wei Zhao |
| BA111M[d] | 2016-01-18 | 43 | VLBA | 1052 | 2048 | Tao An |
| BZ081A[d] | 2021-04-22 | 43 | VLBA | 5551 | 2048 | |
| a2301[d] | 2023-03-29 | 43 | EAVN | 28,800 | 1024 | Wei Zhao |

**Notes.** In chronological order of frequency from lowest to highest. The columns are as follows: (1) project code, (2) date of observation, (3) observing frequency band, (4) participating stations (stations not involved are indicated with a minus sign), (5) total observation time of the target source, (6) data rate, and (7) principal investigator of this observation proposal.
[a] Data from the Astrogeo VLBI FITS image database.
[b] Data from the MOJAVE database (FITS image file).
[c] Correlated data from the NRAO archive.
[d] Our own observational data.

### 2.3. Archival VLBA Data

In order to clarify the nature of the off-axis component and the evolution of the jet orientation, we also analyzed the archival VLBA data of 2021+317 at 15 GHz and 22 GHz. We used six epochs of 22 GHz data from the National Radio Astronomy Observatory (NRAO) archive[10] over the period from 2013 to 2019. The data were selected with a separation of about 1 yr and giving priority to longer observed time and more antennas participating. We also analyzed five epochs of 15 GHz data from the Astrogeo VLBI FITS image database[11] and one epoch from the MOJAVE database (M. L. Lister et al. 2018). These data are in FITS image format and cover the time range from 2021 to 2024. The detailed information of these data is listed in Table 2.

### 2.4. Data Reduction

Following the standard procedures, we used the AIPS software package (E. W. Greisen 2003) for the data calibration and fringe fitting. We used the measured autocorrelation spectra to correct the visibility amplitude errors in the cross-correlation spectra caused by sampler threshold errors. We also performed parallactic angle correction and bandpass calibration. Afterward, we conducted two rounds of global fringe fitting and then applied the corrected results to 2021+317. After completing the above calibrations, we used the CLEAN and self-calibration procedures (including phase and amplitude) in the DIFMAP software package (M. C. Shepherd et al. 1994) to process the data. CLEAN and self-calibration were performed alternately in imaging cycles, resulting in the final image. Data in the Astrogeo VLBI FITS image database and MOJAVE database are fully self-calibrated. Therefore, we directly imaged the data from six epochs in the database using DIFMAP following the above method.

---
[10] https://data.nrao.edu/portal/#/
[11] https://astrogeo.org/vlbi_images/

To model the source structure, we used the MODELFIT task in DIFMAP. According to S. Britzen et al. (2017), while elliptical components may have advantages in modeling data over certain time periods, circular component modeling allows for more consistent analysis across all time periods and more reliable identification of the long-term movement and evolution of individual components. Therefore, we used circular Gaussian models in all our model fittings. The fitted components were cross-identified across the 14 epochs based on their position, flux density, and size. For the relative uncertainties, we followed the method of E. B. Fomalont (1999). The detailed parameters obtained from model fitting are listed in Tables A1, A2, and A3 in the Appendix.

We used the VIMAP program[12] to obtain the spectral index distribution map of the source at 22–43 GHz. To avoid errors caused by differences in uv-sampling and resolution, the image data used in the spectral analysis were pregenerated within a common uv-range, and the restoring beam size was set to the value of the lower-frequency data. Here, we set the pixel scale to 1/20 of the beam size. Since the position of the AGN core changes with frequency, we mask the core region and use the optically thin jet regions to align the images (J.-Y. Kim & S. Trippe 2014). For the data set we used, the uv-range was 37–136.1 $M\lambda$, and the restoring beam size was set to $2.46 \times 1.35$ mas at a position angle of $-20°.8$.

## 3. Observational Results

### 3.1. Source Morphology

Figures 2 and 3 show the naturally weighted EAVN and VLBA observation images of 2021+317, respectively. Table 3 lists the parameters of the synthesized images for all observations. For some epochs, the theoretical image noise is higher than the actual image noise. This discrepancy may be attributed to the highly compact nature of the target source, effective data selection and processing, and an efficient self-calibration procedure, resulting in residual noise in the images being predominantly thermal, with minimal contributions from calibration errors. Moreover, the actual image noise was measured in signal-free regions far from the source structure. In addition, during the observations, weather conditions and system performance were better than the conservative assumptions adopted in the theoretical rms estimates. As noted in standard VLBI practice (J. Zensus et al. 1993; M. L. Lister et al. 2009a), under good observing conditions and with efficient self-calibration, the actual image noise level can be lower than the theoretical thermal noise estimates. The source exhibits a consistent morphology at all frequencies, displaying a compact core region and a southward jet. Transverse elongation was detected in the core region across all epochs. In the 15 GHz images, we can also see faint diffuse emission to the southwest of the core.

Next, we performed model fitting on the VLBA 15, 22, and 43 GHz data using the MODELFIT task in DIFMAP. A circular Gaussian model was used for all data sets. The results indicate that the core region consists of two components located in the southeast–northwest direction, labeled as E and W. The position angle of W relative to E changes slightly across different observation epochs, with the average and standard deviation being $-51°.49$ and $12°.43$, respectively. The bottom panels of Figure 3 show the spectral index distribution maps and the corresponding error maps obtained from the EAVN observations at 22 and 43 GHz. We then obtained the spectral index variation curves along the component E and W directions, with the results shown in Figure 4. Although the spectral index of E is flatter, the spectral index of W does not exceed $-0.5$, so this result is not sufficient to determine which of the two components, E or W, is the core. Both possibilities are considered in the later sections of this paper. The MOJAVE 15 GHz survey continuously monitored the source from 1995 to 2013 and found that since 2010 September, the core region of the source can be fitted with two components (M. L. Lister et al. 2019), which likely correspond to the E and W components fitted in our analysis. The transverse elongation of the core region is relatively common, as seen in objects like Mrk 501 (S. Koyama et al. 2016) and OJ 287 (J. A. Hodgson et al. 2017), and is typically interpreted as a standing shock offset from the inner jet axis or an abrupt jet bending near the core region (Z. R. Weaver et al. 2022).

The jet of the source can consist of multiple components, labeled S1–S5. Among them, components S4 and S5 newly appeared in 2021. By comparing these images, we found that the jet orientation exhibited significant counterclockwise rotation.

### 3.2. Jet Kinematics and Geometry

To better illustrate the motion trajectory of the jet components of 2021+317, we supplemented the data with six epochs from the MOJAVE 15 GHz observations from 2010 to 2013 (M. L. Lister et al. 2019), directly using the model fitting parameters provided in their paper. We plotted the positions of all components of this source across all epochs in Figure 5. In the left panel, we consider E as the core, while in the right panel, we consider W as the core. We can see that the motion of W relative to E does not show an obvious pattern. When considering E as the core, S1 appears to move linearly southward, while the other jet components move southwestward and then turn eastward at a distance of approximately 2 mas from the core, continuing along the southeast direction. From Figure 2, it can also be seen that there is a wide range of diffuse emission in the region where the direction of the jet components changes, possibly due to the interaction between the jet and some form of medium, as suspected, e.g., in the jet in the radio galaxy 4C 41.17 (L. I. Gurvits et al. 1997). The medium responsible for the interaction could be stars within a dense nuclear star cluster, molecular clouds, or star clusters (T. Alexander 2017) or, alternatively, a second black hole located near the center of the host galaxy (S. Britzen et al. 2024). When considering W as the core, the motion of the jet components appears to follow a helical trajectory.

First, taking E and W as the core alternately, we used a simple one-dimensional radial motion fitting method to calculate the proper motions and apparent velocities of S1–S5 and E and W (excluding S2 and S3). We found that S2 and

---

[12] http://astro.snu.ac.kr/~trippe/VIMAP/vimap.html

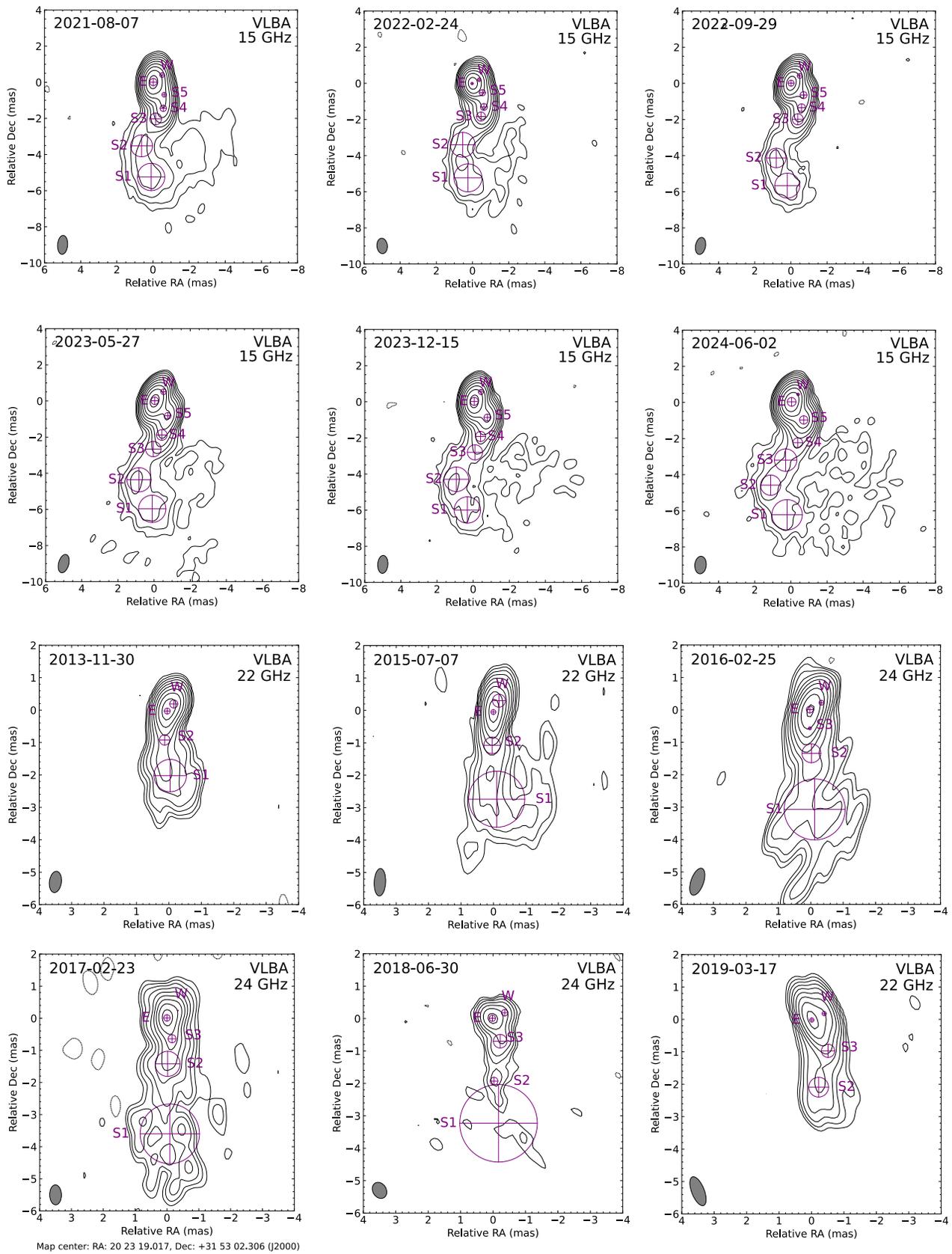

**Figure 2.** The naturally weighted VLBA images of 2021+317 at 15, 22, and 43 GHz. Contours start at 3 times the rms noise level and increase in steps of 2. The gray filled ellipses in the lower left represent the synthesized beam for each image.

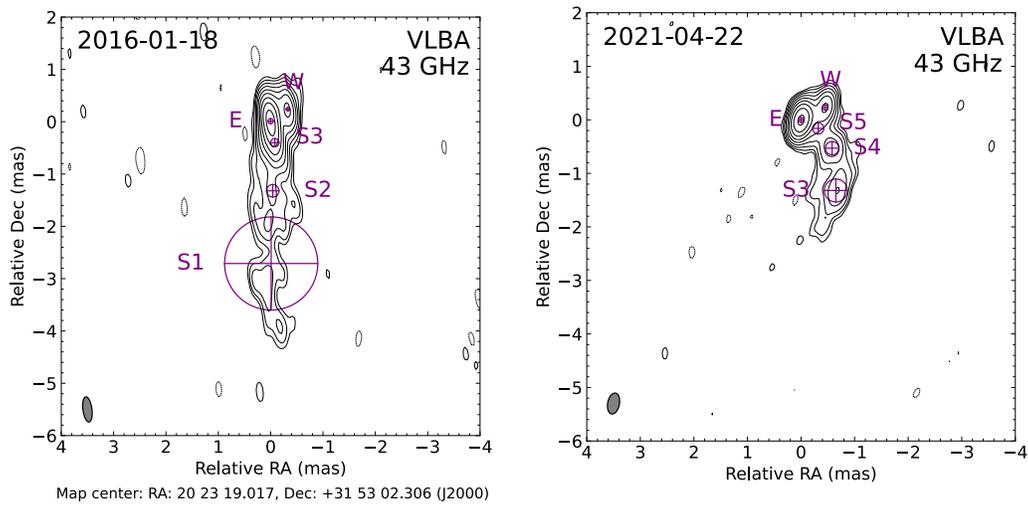

**Figure 2.** (Continued.)

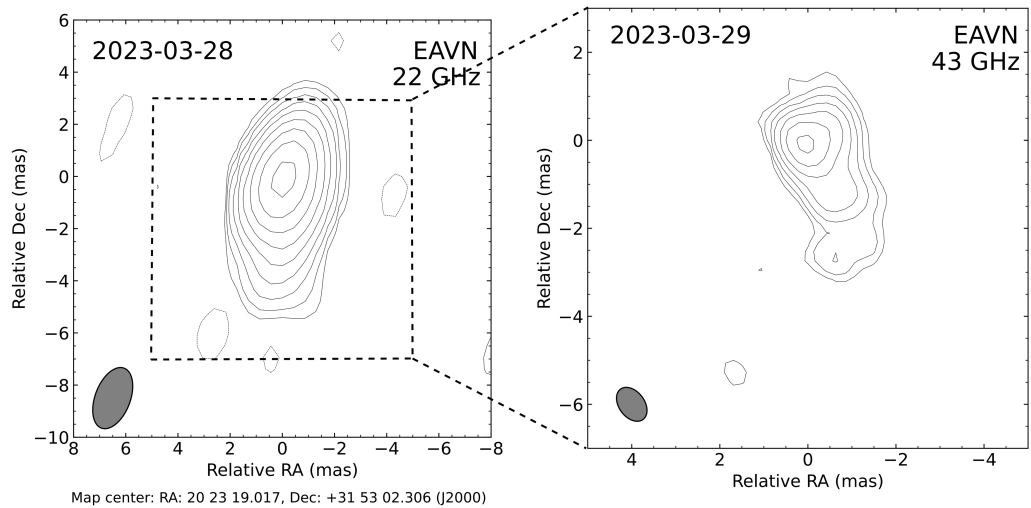

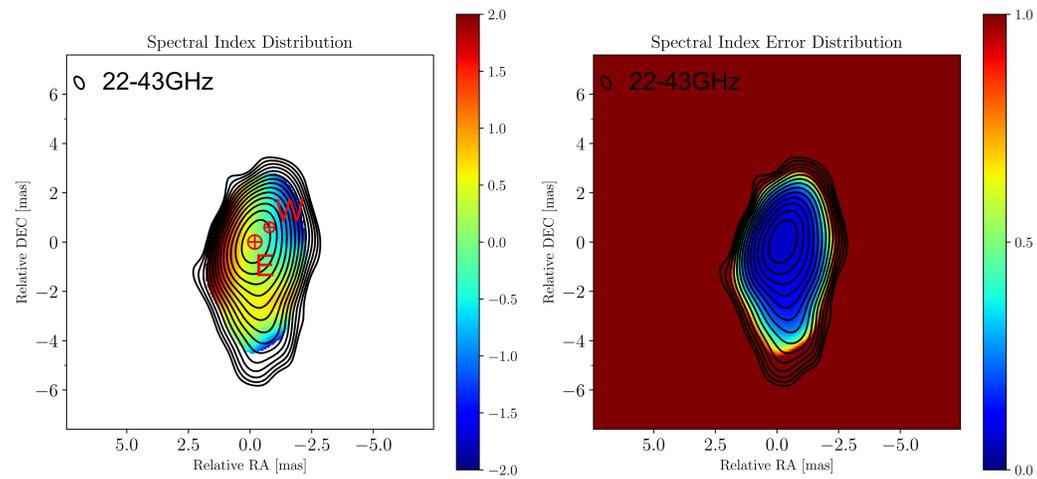

**Figure 3.** Top: the naturally weighted EAVN images of 2021+317 at 22 and 43 GHz. Contours start at 3 times the rms noise level and increase in steps of 2. The gray filled ellipses in the lower left represent the synthesized beam for each image. Bottom left: the spectral index distribution map of 2021+317. The color represents the spectral index $\alpha$ distribution overlaid with the total intensity contours at 22 GHz as in the top left image. Bottom right: the corresponding uncertainties of spectral index map.

**Table 3**
Parameters of Synthesized Images of All Observations (Re)analyzed in the Current Work

| P.C. | Date | Freq. | Array | $\sigma_{\rm rms,th}$ | $\sigma_{\rm rms}$ | $S_{\rm tot}$ | $S_{\rm Peak}$ | Synthesized Beam (Major Axis, Minor Axis, PA) | Figure Reference |
|---|---|---|---|---|---|---|---|---|---|
| | | (GHz) | | (mJy beam$^{-1}$) | (mJy beam$^{-1}$) | (Jy) | (Jy beam$^{-1}$) | (mas, mas, deg) | |
| (1) | (2) | (3) | (4) | (5) | (6) | (7) | (8) | (9) | (10) |
| BL286AA | 2021-08-07 | 15 | VLBA | 0.10 | 0.22 | 1.13 | 0.611 | 1.05, 0.55, −4.3 | |
| BL286AH | 2022-02-24 | 15 | VLBA, −KP | 0.11 | 0.26 | 1.20 | 0.753 | 0.85, 0.58, 5.6 | |
| BL286AO | 2022-09-29 | 15 | VLBA, −KP, −PT | 0.13 | 0.23 | 1.04 | 0.622 | 0.95, 0.55, −10.9 | |
| BL286AU | 2023-05-27 | 15 | VLBA | 0.10 | 0.18 | 0.79 | 0.431 | 1.04, 0.56, −15.3 | Figure 2 |
| BL286BB | 2023-12-15 | 15 | VLBA | 0.10 | 0.19 | 0.89 | 0.470 | 0.99, 0.56, −3.9 | |
| BL286BG | 2024-06-02 | 15 | VLBA | 0.10 | 0.15 | 0.82 | 0.433 | 0.97, 0.63, −4.96 | |
| BR145YB | 2013-11-30 | 22 | VLBA, −FD, −PT | 0.92 | 0.46 | 1.20 | 0.575 | 0.66, 0.37, −7.47 | |
| BM421 | 2015-07-07 | 22 | VLBA | 0.25 | 0.54 | 1.17 | 0.694 | 0.85, 0.35, −2.76 | |
| BR210CD | 2016-02-25 | 24 | VLBA, −SC | 0.87 | 0.39 | 1.24 | 0.677 | 0.87, 0.39, −19.3 | |
| UD001C | 2017-02-23 | 24 | VLBA | 0.22 | 0.31 | 0.84 | 0.34 | 0.62, 0.37, 1.35 | Figure 2 |
| UD001W | 2018-06-30 | 24 | VLBA, −NL | 0.44 | 0.95 | 0.75 | 0.379 | 0.56, 0.44, 31.6 | |
| BZ076A | 2019-03-17 | 22 | VLBA | 1.10 | 0.43 | 0.56 | 0.32 | 0.95, 0.39, 21.5 | |
| a2301 | 2023-03-28 | 22 | EAVN | 0.20 | 0.40 | 0.64 | 0.49 | 2.46, 1.35, −20.8 | Figure 3 |
| BA111M | 2016-01-18 | 43 | VLBA | 0.24 | 0.25 | 0.69 | 0.293 | 0.45, 0.17, 7.54 | |
| BZ081A | 2021-04-22 | 43 | VLBA | 0.11 | 0.26 | 0.48 | 0.272 | 0.40, 0.22, −11 | Figure 2 |
| a2301 | 2023-03-29 | 43 | EAVN | 2.10 | 1.75 | 0.88 | 0.39 | 0.86, 0.58, 36.7 | Figure 3 |

**Note.** In chronological order of frequency from lowest to highest. The columns are as follows: (1) project code, (2) date of observation, (3) observed frequency, (4) participating stations (stations not involved are indicated with a minus sign), (5) theoretical image noise, (6) noise of the resulting image, (7) total flux density, (8) peak brightness, (9) synthesized beam from naturally weighting, and (10) reference to the figure in the paper where the image corresponding to each observational epoch is shown.

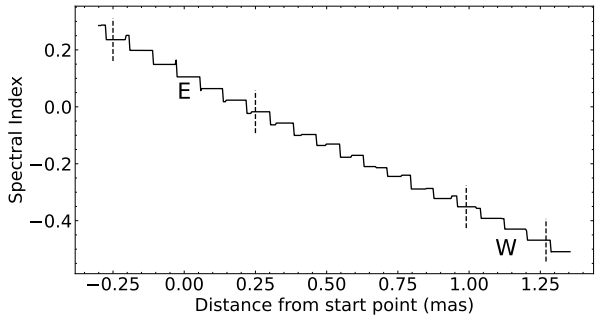

**Figure 4.** The pixel spectral index variation curve of 2021+317 along the E and W directions. The two vertical dashed lines on the left represent the range of E, while the two dashed lines on the right represent the range of W. Due to the small distance between E and W, the number of pixels is limited, resulting in a steplike curve in the slice.

S3 exhibit accelerated motion, so we performed a piecewise simple one-dimensional radial motion fit (B. Rani et al. 2015) and a constant acceleration fit for their accelerated motion. The fitting results are shown in Figure 6, and the fitted parameter values are listed in Table 4. Since the differences in the fitting results are very small, we only present the results with E as the core in the figure. We can observe that the relative motion velocity between the E and W components is very small and can be approximately regarded as nearly stationary.

Next, using Equations (1) and (2) from X. Li et al. (2018), we calculated the intrinsic Lorentz factor ($\Gamma$) of the jet and the angle between the jet and the observer line of sight ($\theta$) for both cases where E and W are considered as the core. The results are $\Gamma \geqslant 9$, $\theta \leqslant 12°.8$ and $\Gamma \geqslant 12.1$, $\theta \leqslant 9°.5$, respectively.

Since we observed in Figure 5 that, with E as the core, the motion directions of all jet components, except for component S1, had significantly changed, we performed two-dimensional sky plane vector fits for all jet components before and after the change in motion direction. The fitting results are listed in Tables 5 and 6, respectively. With W as the core, the motion trajectories of the jet components appear more complex, and we will discuss them in detail later in the text. In this process, we measured the position of each jet component relative to the core across all epochs, assuming in all cases that the bright core feature is stationary (M. L. Lister et al. 2019). Since there is only one epoch of observational data for S4 before the change in its motion direction, we did not fit its vector before the change in direction. From the fitting results, we can see that the change in vector direction for S1 before and after is small, only 14°.2, but the vector direction changes for S5, S3, and S2 are significant, at 57°.4, 70°, and 37.4°, respectively.

Referring to X. Li et al. (2018), we used linear least-squares fitting to determine the relationship between the size of each jet component and its distance from the core, thereby obtaining the projected opening angles of the jet ($\alpha_{proj}$) when using E and W as the core, which are 15°.5 and 14°.5, respectively. We then calculated the intrinsic opening angles of the jet ($\alpha_{int}$) to be $\alpha_{int} \leqslant 3°.5$ and $\alpha_{int} \leqslant 2°.4$, respectively.

### 3.3. Jet Total Intensity Ridgeline

In previous studies, there were various definitions of the AGN jet ridgeline (H. Ro et al. 2023), such as those of M. Perucho et al. (2012), C. M. Fromm et al. (2013a), M. H. Cohen et al. (2015), S. Britzen et al. (2017), and A. B. Pushkarev et al. (2017). To better visualize the possible precession behavior of the jet when using E as the core, we performed model fitting on 2021+317 and obtained the jet ridgeline by connecting each component of the jet. The result is shown in Figure 7. Different colors represent different observed epochs. The color from dark to light indicates that the observed data are becoming more recent.

It can be seen that the ridgeline position angle of the jet exhibits a clear counterclockwise rotation from 2013 to 2021 and a clockwise rotation from 2021 to 2024. However, this may also be due to the fact that we have not yet observed components newer than S5. Initially, S5 moved southwestward, but later, like other jet components except for S1, its direction changed to southeastward, making it appear that the jet has started rotating clockwise. Therefore, to determine the precession period of the source, further observations are needed, possibly even regular observations.

### 3.4. Brightness Temperature Measurements

The brightness temperature can be used to estimate major energy-loss mechanisms and the evolution of intrinsic properties such as jet velocity, particle density, and magnetic field (e.g., M. Kadler et al. 2004; F. K. Schinzel et al. 2012). By fitting the core brightness distribution in the $uv$-plane with a circular Gaussian component via the modelfit procedure in DIFMAP, we can determine the observed rest-frame core brightness temperature (S. G. Jorstad et al. 2005):

$$T_{B,\text{vlbi}}^{\text{obs}} = 1.22 \times 10^{12} \frac{S_\nu(1+z)}{\nu^2 ab} \text{ K.} \quad (1)$$

The intrinsic brightness temperature and Doppler factor can be obtained from the following formulae:

$$T_{B,\text{vlbi}}^{\text{obs}} = T_B^{\text{int}} \times \delta, \quad (2)$$

$$\delta = [\gamma(1-\beta\cos\theta)]^{-1}, \quad (3)$$

where $S_\nu$ corresponds to the fitted core flux density (Jy) for a given observing frequency $\nu$ (GHz), $a$ and $b$ are the FWHM (mas) of the major and minor axes of the elliptical Gaussian core component, and $z$ is the redshift.

From this, we obtained the brightness temperatures of the E and W components detected at 43 GHz in experiment BZ081A: $T_{B,\text{vlbi}}^{\text{obs}} = (3.5 \pm 0.6) \times 10^{10}$ K and $T_{B,\text{vlbi}}^{\text{obs}} = (7.1 \pm 1.7) \times 10^9$ K, respectively. According to the results in Section 3.2, we obtained Doppler factors of $\delta = 3.63$ and $\delta = 4.85$, respectively. And the corresponding intrinsic brightness temperature is $T_B^{\text{int}} = (9.5 \pm 1.7) \times 10^9$ K and $T_B^{\text{int}} = (1.5 \pm 0.4) \times 10^9$ K, respectively. Using the data of MOJAVE 15 GHz from 1995 to 2013, D. C. Homan et al. (2021) calculated the intrinsic peak brightness temperature of the core region as $1.2 \times 10^{10} \sim 8.8 \times 10^{10}$ K, which is larger than our result. Our values are lower than the inverse Compton limit of brightness temperature, $\sim 10^{11.5}$ K (K. I. Kellermann & I. I. K. Pauliny-Toth 1969). They are also lower, although not drastically, than the equipartition limit $5 \times 10^{10}$ K, (A. C. S. Readhead 1994). A similar physical scenario has been found in many sources, such as M87 (J. Y. Kim et al. 2018).

We can get the relationship between magnetic field energy density ($u_B$) and nonthermal particles energy density ($u_p$) using the following formula (A. C. S. Readhead 1994):

$$\frac{u_p}{u_B} = \left(\frac{T'_{\text{eq}}}{T_{B,\text{vlbi}}^{\text{obs}}}\right)^{-\frac{17}{2}}, \quad (4)$$

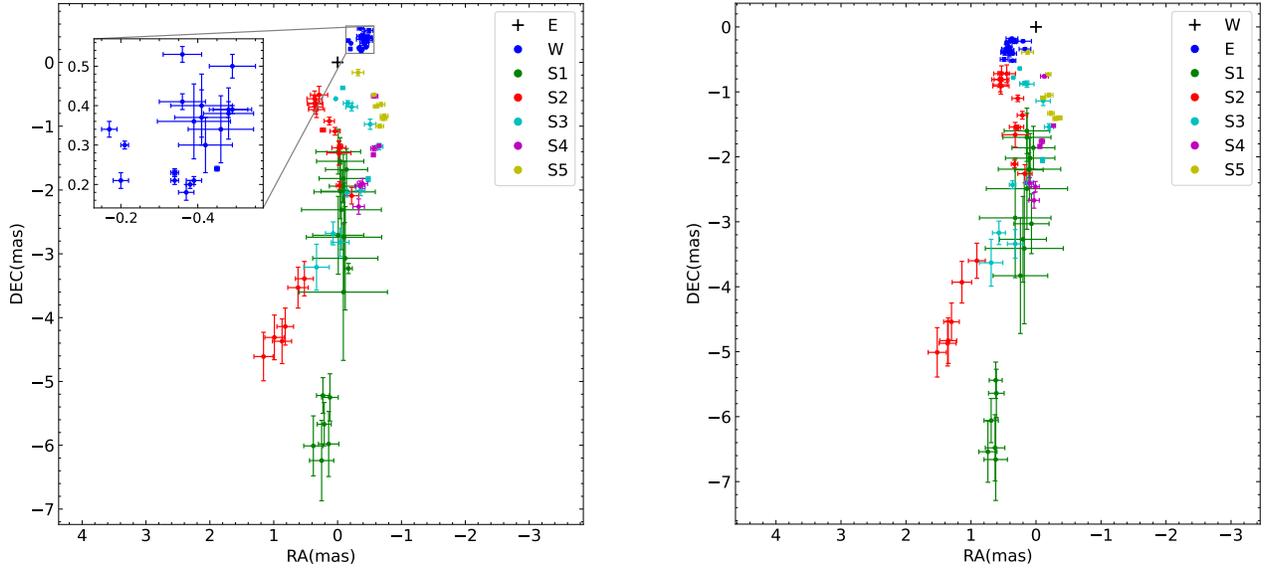

**Figure 5.** The motion trajectories of each jet component from 2010 to 2024 are shown in the figure, with the left panel using the E component as the core and the right panel using the W component as the core. Different components are represented by different colors, and the corresponding relationships are shown in the figure.

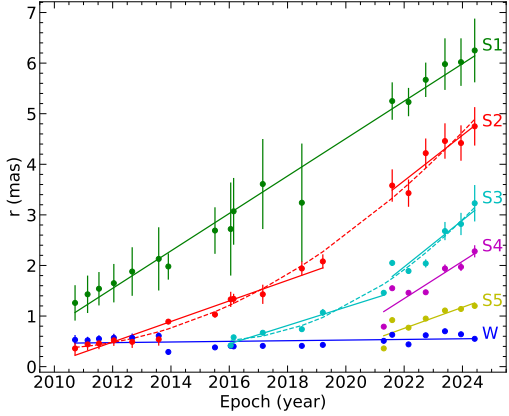

**Figure 6.** The figure shows the evolution of the radial distance to component E. The solid lines represent the linear fit, while the dashed lines represent the accelerated motion fit. Components S2 and S3 were fitted in segments. Different components are represented by different colors, and the corresponding relationships are shown in the figure.

where $T'_{eq} = \delta \times T_{eq}$. Therefore, for E and W, we can determine $\lg\left(\frac{u_p}{u_B}\right) = -6.1$ and $\lg\left(\frac{u_p}{u_B}\right) = -13$, respectively. That is, when E is the core of the source, the magnetic field energy density in the core region is $10^{6.1}$ times greater than the energy density of nonthermal particles. When W is the core of the source, the magnetic field energy density in the core region is $10^{13}$ times greater than the energy density of nonthermal particles.

## 4. Discussion

### 4.1. Properties of the E and W Components

VLBI monitoring enables us to study the variable kinematics of the jet components. From Figure 5, it can be seen that the relative position between the W and E components is relatively stable, showing no obvious movement trend, while other jet components exhibit significant motion.

Stationary features are a common characteristic in AGN jets (e.g., S. G. Jorstad et al. 2001; S. Britzen et al. 2010; C. M. Fromm et al. 2013b). AGNs are often accompanied by high-speed jets ejected from their core regions, and when these jets interact with the surrounding galactic medium (such as intergalactic or galaxy cluster gas), they can form a shock wave at the front of the jet, which may become relatively stable if the jet is sustained and stable, and the density of the external medium is relatively uniform. This process leads to the formation of a stationary jet feature (e.g., R. D. Blandford & D. G. Payne 1982; M. C. Begelman et al. 1984; A. C. Fabian 2012). Specifically, in a straight jet, the stationary features can be generated by recollimation shocks, magnetohydrodynamic and/or Kelvin–Helmholtz instabilities, or magnetic pinches (e.g., D. L. Meier et al. 2001; P. E. Hardee 2006; A. P. Marscher 2009; R. K. Joshi & I. Chattopadhyay 2023). According to numerical simulations by J. L. Gómez et al. (1997), a recollimated shock can be affected by a passing disturbance that causes its position to fluctuate. To determine this, VLBI multiepoch phase reference observations are required (J. A. Hodgson et al. 2017). The bending of the jet also produces an apparent quasi-stationary feature, where the viewing angle of the jet might become smaller after the bend, thus leading to an increased Doppler boosting factor. Yet another scenario involves a formation of a shock wave that deflects the flow (A. Alberdi et al. 1993).

Whether E or W is the core component, the other component may be a stationary shock. The jet originates in the core component and deflects after reaching the stationary shock, thus defining the observed shape of the jet. It has been suggested in some studies (e.g., F. D. D'Arcangelo et al. 2007; A. P. Marscher et al. 2008) that a stationary shock in an unresolved core of the AGN jet may be responsible for the persistent high-level polarization in blazars.

Yet another alternative possibility is that E is the core, while W could be a counterjet. Such a configuration might be caused by various factors. For example, the interstellar medium surrounding the AGN may not be uniformly distributed, and the jet might be influenced by the environment along the propagation path, leading to bending or deflection

**Table 4**
Physical Parameters of the Radio Emission

| ID | E as the Core | | | W as the Core | | |
|---|---|---|---|---|---|---|
| | $\mu$ (mas yr$^{-1}$) | $\beta_{app}$ (c) | $a$ (mas yr$^{-2}$) | $\mu$ (mas yr$^{-1}$) | $\beta_{app}$ (c) | $a$ (mas yr$^{-2}$) |
| (1) | (2) | (3) | (4) | (5) | (6) | (7) |
| E | ⋯ | ⋯ | ⋯ | 0.006 ± 0.005 | 0.14 ± 0.12 | ⋯ |
| W | 0.006 ± 0.005 | 0.14 ± 0.12 | ⋯ | ⋯ | ⋯ | ⋯ |
| S1 | 0.37 ± 0.01 | 8.92 ± 0.24 | ⋯ | 0.38 ± 0.02 | 9.16 ± 0.48 | ⋯ |
| S2$_1$ | 0.2 ± 0.01 | 4.82 ± 0.24 | 0.04 ± 0.01 | 0.16 ± 0.01 | 3.86 ± 0.24 | 0.06 ± 0.01 |
| S2$_2$ | 0.45 ± 0.09 | 10.85 ± 2.17 | ⋯ | 0.50 ± 0.13 | 12.05 ± 3.13 | ⋯ |
| S3$_1$ | 0.18 ± 0.02 | 4.34 ± 0.48 | 0.06 ± 0.02 | 0.14 ± 0.03 | 3.37 ± 0.72 | 0.1 ± 0.02 |
| S3$_2$ | 0.46 ± 0.09 | 11.09 ± 2.17 | ⋯ | 0.54 ± 0.13 | 13.01 ± 3.13 | ⋯ |
| S4 | 0.37 ± 0.08 | 8.92 ± 1.93 | ⋯ | 0.5 ± 0.11 | 12.05 ± 2.65 | ⋯ |
| S5 | 0.21 ± 0.06 | 5.06 ± 1.45 | ⋯ | 0.28 ± 0.08 | 6.75 ± 1.93 | ⋯ |

**Note.** The subscripts of S2 and S3 indicate that they were divided into segments for fitting, with the first and second segments represented by subscripts 1 and 2, respectively. Column (1): component label. Columns (2)–(4): proper motion, apparent velocity, and acceleration obtained from fitting the overall accelerated motion of this component (without segmented fitting) when E is considered as the core, respectively. Columns (5)–(7): same as columns (2)–(4) but with W considered as the core.

**Table 5**
Two-dimensional Sky Plane Vector Fit to the Jet Components When Using E as the Core (before the Change in Motion Direction)

| ID | No. | ⟨S⟩ (mJy) | ⟨R⟩ (mas) | ⟨d$_{proj}$⟩ (pc) | ⟨θ⟩ (deg) | φ (deg) | |θ − φ| (deg) |
|---|---|---|---|---|---|---|---|
| (1) | (2) | (3) | (4) | (5) | (6) | (7) | (8) |
| S5 | 4 | 82 | 0.75 | 3.72 | −130.3 | −147.2 | 16.9 |
| S3 | 6 | 116 | 0.82 | 4.07 | −163.8 | −140.2 | 23.6 |
| S2 | 14 | 149 | 0.98 | 4.86 | 174.3 | −169 | 16.7 |
| S1 | 13 | 200 | 2.23 | 11.06 | −173.4 | 166.9 | 19.7 |

**Note.** The columns are as follows: (1) component label, (2) number of epoch fits, (3) average flow density, (4) average distance of jet components from the core, (5) average projected distance of jet components from the core, (6) average position angle of the jet component with respect to the core, (7) velocity vector position angle, and (8) shift between the position angle of the velocity vector and the mean position angle.

**Table 6**
Two-dimensional Sky Plane Vector Fit to the Jet Components When Using E as the Core (after the Change in Motion Direction)

| ID | No. | ⟨S⟩ (mJy) | ⟨R⟩ (mas) | ⟨d$_{proj}$⟩ (pc) | ⟨θ⟩ (deg) | φ (deg) | |θ − φ| (deg) |
|---|---|---|---|---|---|---|---|
| (1) | (2) | (3) | (4) | (5) | (6) | (7) | (8) |
| S5 | 3 | 123 | 1.15 | 5.70 | −142.1 | 155.4 | 62.5 |
| S4 | 6 | 45 | 1.78 | 8.83 | −163.2 | 160.1 | 36.7 |
| S3 | 6 | 22 | 2.45 | 12.16 | −176.3 | 150.2 | 33.5 |
| S2 | 6 | 24 | 4.15 | 20.56 | 168.6 | 153.6 | 15 |
| S1 | 6 | 29 | 5.73 | 28.44 | 177.8 | 152.7 | 25.1 |

**Note.** The columns are as follows: (1) component label, (2) number of epoch fits, (3) average flow density, (4) average distance of jet components from the core, (5) average projected distance of jet components from the core, (6) average position angle of the jet component with respect to the core, (7) velocity vector position angle, and (8) shift between the position angle of the velocity vector and the mean position angle.

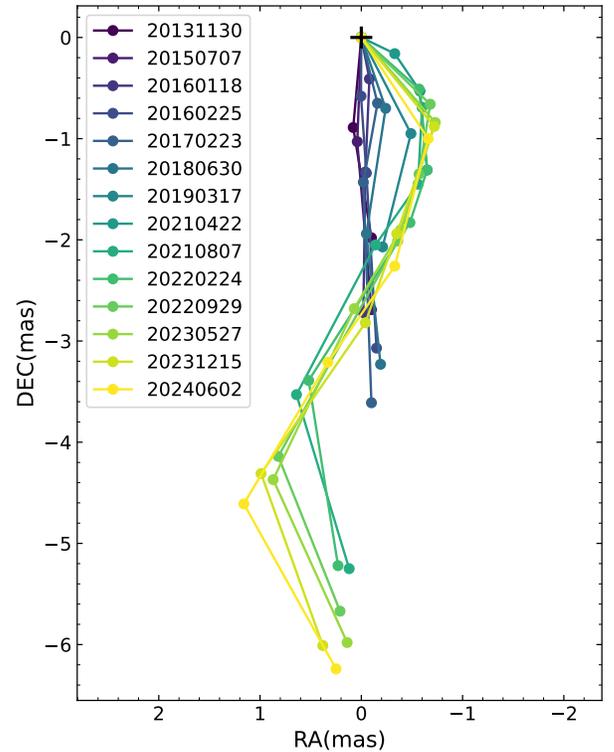

**Figure 7.** The jet total intensity ridgeline is shown. Different colors represent different observed epochs. The color from dark to light indicates that the observed data are becoming more recent.

(G. V. Bicknell 1995; L. I. Gurvits et al. 1997). Instabilities within the jet, such as Kelvin–Helmholtz instability or helical structures, also might cause fluctuations or deviations in the jet path, making it appear misaligned with the counterjet (P. E. Hardee 2000). Additionally, if there is a certain inclination angle between the spin axis of the AGN central black hole and the angular momentum axis of the accretion disk, the initial directions of the jet and counterjet might be slightly different (D. Garofalo et al. 2010).

### 4.2. Jet Wobbling

From Figure 5, it can be seen that when E is considered as the core, all jet components except W seem to have a systematic change of the jet components' position angle in the counterclockwise direction. When W is considered as the core, the jet appears to have multiple bends, showing a helical trajectory. For the first case, the pattern is more evident in Figure 7. The results show that the jet of this source has obvious counterclockwise rotation of its position angle during the observing period, but a periodicity of position angle variation is not visible in our data. To prove any hypothetical periodicity, multiple periods over a longer observing period are needed (R. Lico et al. 2020); the covered observing time interval might be insufficient for detecting a periodicity. Periodicity in jet wobbling is very rare and is often interpreted as evidence for the presence of a binary black hole system, as suggested by Z. R. Weaver et al. (2022). A similar interpretation was proposed by M. L. Lister et al. (2003) for the source 4C+12.56.

There are various reasons for AGN jet oscillation; for example, the dynamic changes in the magnetic field can lead to instability, which may cause the jet to oscillate. The magnetic field plays a key role in the formation and maintenance of the jet, and any change in the magnetic field may lead to a change in the jet path (N. Steinle et al. 2024). The jet may be deflected or wobbled by its interaction with the galactic or intergalactic medium as it passes through. The inhomogeneity of the medium, such as density fluctuation, can affect the trajectory of the jet (E. Kun et al. 2014). The direction of the jet may be influenced by the rotation of the central black hole and the dynamics of the accretion disk. In particular, if the spin axis of the black hole is inconsistent with the axis of the jet, this inconsistency can change over time, causing the jet to wobble (K. Hada 2019). There are other factors that can cause the wobble of the jet to appear periodic, such as jet nozzle precession, rotation, or spiral jet structure, where the observed variation is due to a changing viewing angle, and therefore the Doppler factor also varies (R. Lico et al. 2020). Another possible cause is the Bardeen–Peterson effect, in which the angular momentum of the outer accretion disk is misaligned with that of the black hole. As matter gradually accretes, the inner disk aligns with the black hole's spin axis (J. M. Bardeen & J. A. Petterson 1975; S. Liu & F. Melia 2002; A. Caproni et al. 2004). However, C. Nixon & A. King (2013) argued that due to the combined effects of Lense–Thirring precession and accretion disk viscosity, it is unlikely that any type of black hole system alone can fully account for the observed jet precession. In general, the angular momentum of the disk is not large enough to cause jet precession on short timescales compared to the angular momentum of the rotating black hole. A. Caproni et al. (2006) argued that other precession mechanisms cannot be easily ruled out due to uncertainties in many observations. In addition, there may be more than one mechanism at work in a given source.

Numerous jet periodic oscillation models have been proposed, including those by A. Caproni et al. (2009), T. An et al. (2010), S.-J. Qian (2011), P. Mohan et al. (2016), S. Britzen et al. (2019), and M. S. Butuzova & A. B. Pushkarev (2020). These models include both the jet precession model and the spiral jet model. In the jet precession model, it is assumed that the apparent superluminal knots are emitted from a jet nozzle that precesses around a fixed jet axis. After being ejected, the knots move with a constant phase, possibly along a straight line (A. Caproni et al. 2009) or initially along a parabolic path for a period of time, and then move in a straight line; in this case, the motion trajectories of different jet components should be parallel straight lines (S.-J. Qian 2011). The spiral jet model holds that the jet is a continuous flow with a spiral shape, and the spiral jet originates from the Kelvin–Helmholtz instability (M. S. Butuzova & A. B. Pushkarev 2020). Furthermore, the existence of a poloidal (jet-parallel) magnetic field in the inner region of the jet may contribute to the development of a helical geometry (J. Röder et al. 2025). In S. N. Molina et al. (2014), the angle between the jet and the line of sight should be small enough, meaning that the line of sight is contained within the opening angle of the jet. When the front of the jet is seen, the innermost jet emission region rotates around the jet axis. These tracks may be generated by helical or quasi-helical magnetic fields passing through the magnetic dominant region of the innermost jet. The jet feature we actually observed should revolve around a fixed point, that is, around the actual jet axis projected on the sky plane. Or, for the general case, where the angle of the jet does not include the line of sight inside, we will see that the jet appears to have a very obvious curved trajectory (M. S. Butuzova & A. B. Pushkarev 2020). H. Falcke et al. (1996) suggests that helical structures associated with radio jets appear to be common, as some VLBI components of quasar jets appear to follow helical trajectories (W. Steffen et al. 1995), and $H_\alpha$ jets in NGC 4258 consist of helically twisted triple helices (G. Cecil et al. 1992).

For 2021+317, when E is considered as the core, the jet components appear to propagate in a fixed direction rather than following a fixed helical trajectory, which is more consistent with the description of the jet precession model. However, when W is considered as the core, the jet seems to propagate along a helical trajectory. Therefore, we performed fitting for both cases. First, we used the jet precession model proposed by S.-J. Qian (2011) to fit the motion trajectories of the jet components for this source when considering E as the core. The fitting results are shown in the left panel of Figure 8. Since the jet components of this source seem to have encountered some medium during propagation that changed their direction of motion, we only used the positions before the change in their motion direction for this fitting. We obtained the precession phases for S1–S5 (except S4) through fitting; they are 0.0004, 0.006, 1.79, and 3.14 rad, respectively. If we directly backtrack based on the proper-motion fitting results in Section 3.2, we can obtain the emission times for S1–S5 (except S4) as 2007.84, 2007.81, 2011.36, and 2017.74, respectively. Then, we can fit the precession period of the jet source to be 19.8 yr.

Then, we used the helical jet model proposed by W. Steffen et al. (1995) to fit the motion trajectories of the jet components for this source when considering W as the core. The fitting results are shown in the right panel of Figure 8. This model regards the discernible structures observed within the jet as "plasma blobs" (emission regions) moving along a three-dimensionally curved trajectory. Under given initial conditions —such as the initial location of the disturbance and the initial angular velocity—the motion of these blobs is assumed to be constrained by the conservation of kinetic energy, the conservation of momentum along the jet axis, and a fixed jet opening angle. We performed a fit to the source using $\mu = 0.5$ mas yr$^{-1}$, $\theta = 9°.5$, and $\alpha_{int} = 2°.4$ and assumed that the initial disturbance occurs very close to the central engine. We found from the fit that the initial disturbance is located at (0 mas, 0.003 mas, 135°.1) in the source comoving cylindrical coordinate system, with the central engine at the origin, and that the initial angular velocity is 15.3 rad yr$^{-1}$.

Although both the precessing jet model and the helical jet model can fit this source morphology, the precessing jet model provides a better explanation for the diffuse emission observed southwest of the core region in Figure 2. In this model, the jet trajectory covers this area, and as the jet moves through, it encounters some form of medium, causing its direction to change and producing the diffuse emission. In contrast, the helical jet model trajectory does not include this region. Therefore, we are more inclined to identify E as the core of the source.

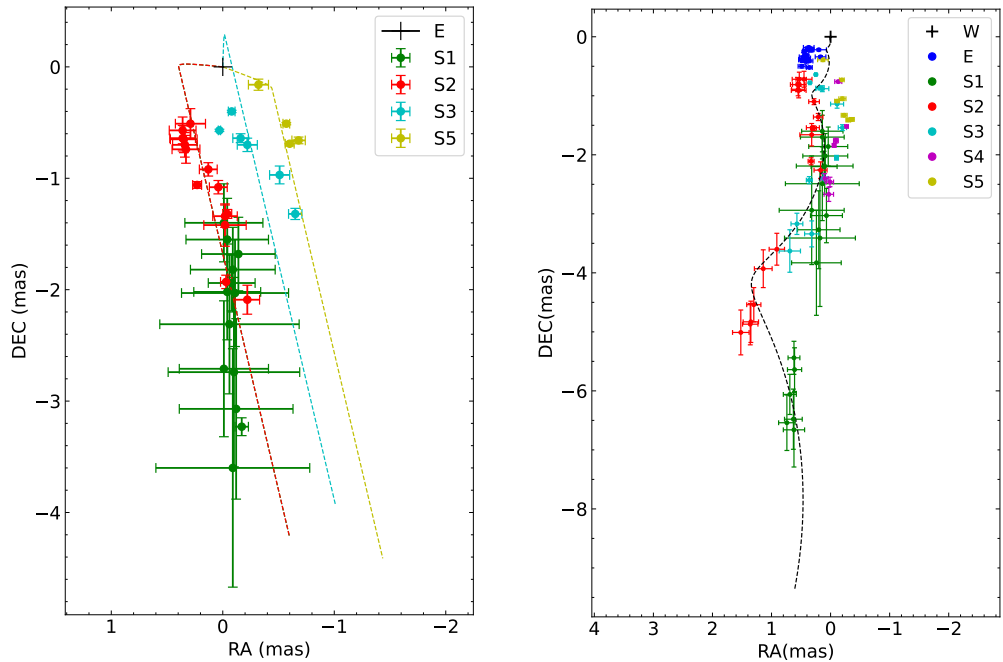

**Figure 8.** The left panel shows the results of fitting the motion trajectories of the jet components in 2021+317 using the jet precession model, taking E as the core. Different jet components are represented by different colored points, and dashed lines in the corresponding colors indicate the trajectories predicted by the model for these components. Because the precession phases of S1 and S2 are very close, their fitted curves in the figure almost overlap and are hard to distinguish. The right panel shows the results of fitting the motion trajectories of the jet components in this source using the helical jet model, taking W as the core. The black dashed lines indicate the trajectories predicted by that model.

### 4.3. Jet Kinematics

From the results in Section 3.2, we can see that when E is taken as the core, all jet components except S1 undergo a significant change in direction after moving to a position approximately 2 mas southwest of the core. They shift from moving southwest to southeast. It seems that there is some dense medium at this location, which interacts with the jet components, causing a change in the direction of motion. This interaction also causes the medium to emit diffuse radiation (A. Alberdi et al. 2000), as we can observe in Figure 2. The average brightness of this diffuse emission is 0.9 mJy beam$^{-1}$, which is approximately 4 times the actual noise level of the corresponding image.

Similar situations have been observed in other sources as well, such as 0858−279 (N. A. Kosogorov et al. 2022), PKS 1717+177 (S. Britzen et al. 2024), and 4C 41.17 (L. I. Gurvits et al. 1997). The bending of AGN jets is a common phenomenon, typically associated with interactions between the jets and the surrounding medium, fluid or magnetohydrodynamic instabilities, or periodic precession of the jet nozzle (T. An et al. 2010). We believe that the change in the direction of motion of the jet components in 2021+317 is due to the interaction between the jets and the surrounding medium. In most galactic nuclei, targets that can interact with jets include stars within dense nuclear star clusters as well as molecular clouds and star clusters (T. Alexander 2017). This interaction may result in a deflection of the jet propagation direction, accompanied by the development of shocks and polarization enhancement (G. F. Paraschos et al. 2024). The deflection is entirely due to hydrodynamic effects, where the collision of the jet with an obstacle creates a shock wave within the jet, converting its kinetic energy into thermal energy, causing the jet to slow down and potentially change direction; consequently, the jet cone becomes broader downstream (S. Britzen et al. 2024). Higgins verified through hydrodynamic simulations that the significant jet deflection is caused by the interaction between the jet and a cloud (S. W. Higgins et al. 1999). It is worth mentioning that L. I. Gurvits et al. (1997) estimated the mass of such a dense clump, but their estimation was based on the assumption that the jet is parallel to the plane of the sky, which does not apply to this source. Another possibility is that the bending of the jet is caused by a second black hole near the center of the host galaxy. During galaxy evolution, mergers can occur, and the final stage of a galaxy merger involves a bound pair of two massive black holes. The second black hole can approach the primary black hole from any angle with a certain probability of being on an orbit that intersects the primary black hole jet. Its gravitational field can cause the jet of the primary black hole to bend, and its external magnetic field component can also interact with the primary black hole jet to induce bending (S. Britzen et al. 2024). In the case of interaction with an ambient cloud, the cloud may be compressed and ionized, leading to free–free absorption (K. I. Kellermann 1966), which would steepen the spectrum below the peak frequency, unlike the case of synchrotron self-absorption (S. Kameno et al. 2000). Therefore, we can verify this hypothesis through subsequent polarization observations. High-resolution circular polarization imaging offers a promising means to distinguish whether the internal magnetic field structure of the jet is purely poloidal, thereby providing a diagnostic for assessing whether the jet may possess a helical geometry (J. A. Kramer & N. R. MacDonald 2021; G. F. Paraschos et al. 2024).

M. L. Lister et al. (2019) used the data from 2006 to 2013 to fit the constant acceleration of S1 and found that S1 exhibited very significant acceleration, with an acceleration of 0.048 mas yr$^{-2}$, but we only found acceleration in S2 and S3. This may be because the data we used are from different time periods. From 2006 to 2013, the S1 component is still accelerating, but the acceleration has gradually decreased, and from 2010 to 2024, the acceleration of S1 has approached zero. In AGNs, the jets accelerate at the beginning of emission. R. D. Blandford & D. G. Payne (1982) believed that in the initial stage of jet formation, a strong magnetic field formed around the accretion disk, and the material would accelerate along the magnetic field line. R. D. Blandford & R. L. Znajek (1977) proposed that the rotating black hole can transfer the rotational energy to the jet through its magnetic field, thus accelerating the jet, and this process depends on the interaction between the black hole and the magnetic field around it. In addition, some studies (e.g., P. A. Becker et al. 2008; S. Mandal & S. K. Chakrabarti 2008) have proposed that standing shocks within the unresolved core of the AGN jet may play a major role in accelerating particles near the base of the jet and may be responsible for the persistent high levels of polarization in the blazars (e.g., F. D. D'Arcangelo et al. 2007; A. P. Marscher et al. 2008). However, when the jet is far away from the core region, its acceleration process will slow down, possibly because as the jet is far away from the black hole and the accretion disk, the surrounding magnetic field strength will gradually weaken, resulting in a decrease in the efficiency of magnetic acceleration (S. S. Komissarov et al. 2009). Meanwhile, during the propagation of the jet, the interaction with the surrounding media will lead to energy dissipation, further leading to a decrease in the acceleration efficiency (A. P. Marscher 2014). When the speed of matter approaches the speed of light, its mass increases due to relativistic effects, resulting in further acceleration requiring more energy, so the acceleration process gradually slows down (A. Tchekhovskoy et al. 2011). The combination of these factors prevents the jet from accelerating after moving away from the core but allows it to propagate at a more stable speed. Another possible scenario is that the jet components propagate at relativistic speeds, and due to the Doppler effect, a change in viewing angle can produce an apparent acceleration. Even if the intrinsic velocity of the jet components remains constant, the relative change with respect to the line of sight can make the jet appear as if it is accelerating (C. M. Urry & P. Padovani 1995). The observed acceleration in S2 and S3 may be due to a change in their direction of motion after interacting with some form of medium, which alters the angle between their motion and our line of sight and thus creates the appearance of acceleration.

## 5. Summary

In our research, we conducted an in-depth observation and analysis of the jet structure in 2021+317 using our own and archival VLBI data at 15, 22, and 43 GHz. Through this work, we explored the internal mechanisms and dynamic behaviors of the AGN jet in the source 2021+317. Here are some of the key findings and analysis results from our study.

1. We obtained images of the source from 2013 to 2024 at 15, 22, and 43 GHz over a total of 16 epochs, with three epochs at 43 GHz and one epoch at 22 GHz being our own observational data. The other data were sourced from the NRAO archive, the Astrogeo VLBI FITS image database, and the MOJAVE database. The images show that the source consists of a compact core region and a jet extending southward. In the 15 GHz images, we also observed diffuse radiation southwest of the core region. Within the core region, we detected two stationary components, E and W, and one of them is likely to be the jet core, while the other one is a stationary knot. If component E is the core, component W is located off the jet axis. Both of these components are clearly visible in all observed epochs.

2. Using the 22 and 43 GHz EAVN data from 2023 March, we obtained the spectral index distribution of the source and further derived its spectral index distribution along the direction of components E and W. The results show that E has a flatter spectral index than W, but since the spectral index of W is not less than $-0.5$, we cannot determine which component is the core based on this criterion alone.

3. By alternately taking the E and W components as the core, we fitted the proper motions and apparent velocities of all the jet components. For the S2 and S3 components, we performed segmented fittings and also applied a constant acceleration model to fit their accelerated motions. Consequently, we obtained the Lorentz factors and viewing angles of the jet for the cases of taking E and W as the core, yielding $\Gamma \geqslant 9, \theta \leqslant 12°.8$ and $\Gamma \geqslant 12.1, \theta \leqslant 9°.5$, respectively. We also obtained the intrinsic opening angles of the jet in the two cases, which are $\alpha_{\rm int} \leqslant 3°.5$ and $\alpha_{\rm int} \leqslant 2°.4$, respectively.

4. Assuming component E as the core and using multiepoch observational data, we identified changes in the total intensity ridgeline of the jet, revealing a significant counterclockwise rotation of its position angle. We performed unaccelerated two-dimensional sky plane vector fitting for the motion trajectories of the jet components and found that the vector direction change for S1 is small, only 14°.2, while the changes for S5, S3, and S2 are more significant, at 57°.4, 70°, and 37°.4, respectively. We speculate that the change in the motion direction of the jet components results from interactions between the jet and the surrounding dense medium. The bending of the jet may be related to hydrodynamic effects, magnetohydrodynamic instabilities, or factors such as a binary black hole system.

5. We calculated the intrinsic brightness temperature and Doppler factor of the source by taking the components E and W as the core, obtaining $T_{B,{\rm vlbi}}^{\rm obs} = (3.5 \pm 0.6) \times 10^{10}$ K, $\delta = 3.63$ and $T_{B,{\rm vlbi}}^{\rm obs} = (7.1 \pm 1.7) \times 10^{9}$ K, $\delta = 4.85$, respectively. In these two cases, the magnetic field energy density in the core is $10^{6.1}$ and $10^{13}$ times the energy density of nonthermal particles, respectively.

6. Based on the observations of jet wobbling, we explored its possible physical mechanisms, including the dynamics of binary black hole systems, jet instabilities, and interactions with the surrounding medium. Assuming E as the core, the jet precession model fit indicates a precession period of 19.8 yr for this source. Assuming W as the core, the helical jet model fit suggests an initial disturbance at (0 mas, 0.003 mas, 135°.1) with an angular velocity of 15.3 rad yr$^{-1}$. Since the precession model better explains the diffuse emission southwest of the core region, while the helical model does not cover this area, we are more inclined to consider E as the core.

In summary, our research, through high-resolution VLBI observations of the AGN 2021+317, provided new insights into the jet physics, dynamical processes, and mechanisms of interaction between the jet and the central supermassive black hole. Through these analyses, we not only revealed the complexity of the jet in the AGN 2021+317 but also offered broad references for understanding the broader phenomena of AGN jets. Future multiepoch monitoring of this source will be essential to trace the kinematics of newly emerging jet components, thereby determining which model better explains the large-scale jet morphology. In addition, higher-frequency and higher-resolution observations will help probe the innermost jet regions and constrain the jet behavior closer to the core, which is critical for identifying the most appropriate model for describing the global jet structure. Furthermore, high-resolution circular polarization imaging may be used to assess whether the internal magnetic field structure is purely poloidal, thereby providing a diagnostic for determining whether the jet exhibits a helical geometry.

## Acknowledgments

We would like to thank the PIs of all the data used in this study, including Matthew Lister, Andreas Brunthaler, James Miller-Jones, Mark Reid, Ingyin Zaw, and VLBA Operations, for providing publicly available data. This research would not have been possible without their support. We sincerely thank Prof. Ken Kellermann for his valuable suggestions and assistance during the preparation of this paper. This research has made use of data from the MOJAVE database that is maintained by the MOJAVE team (M. L. Lister et al. 2018). The Very Long Baseline Array is operated by the National Radio Astronomy Observatory, a facility of the National Science Foundation, operated under cooperative agreement by Associated Universities, Inc. We used in our work the Astrogeo VLBI FITS image database, doi:10.25966/kyy8-yp57, maintained by Leonid Petrov. This work has made use of the East Asian VLBI Network (EAVN), which is operated under cooperative agreement by the National Astronomical Observatory of Japan (NAOJ), the Korea Astronomy and Space Science Institute (KASI), Shanghai Astronomical Observatory (SHAO), Xinjiang Astronomical Observatory (XAO), Yunnan Astronomical Observatory (YNAO), the National Astronomical Research Institute of Thailand (Public Organization) (NARIT), and the National Geographic Information Institute (NGII), with operational support by Ibaraki University (for the operation of the Hitachi 32 m and Takahagi 32 m telescopes), Yamaguchi University (for the operation of the Yamaguchi 32 m telescope), and Kagoshima University (for the operation of the VERA Iriki antenna). This work was supported by the National Key R&D Program of China (2018YFA0404602). L.I.G. and X.Y.H. gratefully acknowledge the Chinese Academy of Sciences PIFI program, grant No. 2024PVA0008.

## Appendix

We performed model fitting for the source using the MODELFIT task in DIFMAP. In all model fittings, we adopted circular Gaussian models. The fitted components were cross-identified across all epochs based on their position, flux density, and size. For relative uncertainties, we followed the method of E. B. Fomalont (1999).

The detailed parameters obtained from model fitting are listed in Tables A1, A2, and A3.

**Table A1**
15 GHz Model Fitting Results

| P.C. (1) | Date (2) | Freq. (GHz) (3) | ID (4) | Flux Density (mJy) (5) | Positions with Respect to E | | d (mas) (8) |
| --- | --- | --- | --- | --- | --- | --- | --- |
| | | | | | $r$ (mas) (6) | PA (deg) (7) | |
| BL286AA | 2021-08-07 | 15 | E  | 813.4 ± 72.1 | 0 ± 0.01    | 0 ± 36.7      | 0.4 ± 0.03  |
|         |            |    | W  | 50.1 ± 29.6  | 0.63 ± 0.01 | −51.5 ± 1.7   | 0.26 ± 0.03 |
|         |            |    | S5 | 82.3 ± 24.9  | 0.92 ± 0.02 | −139 ± 1.3    | 0.25 ± 0.04 |
|         |            |    | S4 | 105.2 ± 25.8 | 1.55 ± 0.03 | −159 ± 1.1    | 0.35 ± 0.06 |
|         |            |    | S3 | 22.8 ± 16.7  | 2.05 ± 0.06 | −176.2 ± 1.8  | 0.65 ± 0.13 |
|         |            |    | S2 | 25.7 ± 15.4  | 3.59 ± 0.32 | 169.7 ± 5.2   | 1.21 ± 0.65 |
|         |            |    | S1 | 50.1 ± 25.9  | 5.25 ± 0.37 | 178.7 ± 4.1   | 1.51 ± 0.75 |
| BL286AH | 2022-02-24 | 15 | E  | 726.2 ± 40.2 | 0 ± 0.01    | 0 ± 6.8       | 0.13 ± 0.01 |
|         |            |    | W  | 171.3 ± 23.3 | 0.44 ± 0.01 | −61.1 ± 0.7   | 0.19 ± 0.01 |
|         |            |    | S5 | 114.6 ± 18.1 | 0.77 ± 0.01 | −131.8 ± 0.9  | 0.34 ± 0.02 |
|         |            |    | S4 | 59.4 ± 11.8  | 1.46 ± 0.02 | −153.4 ± 0.9  | 0.38 ± 0.05 |
|         |            |    | S3 | 38.9 ± 9.7   | 1.89 ± 0.04 | −165.2 ± 1.1  | 0.5 ± 0.07  |
|         |            |    | S2 | 25.8 ± 10.7  | 3.43 ± 0.27 | 171.2 ± 4.5   | 1.37 ± 0.54 |
|         |            |    | S1 | 47.3 ± 17.3  | 5.23 ± 0.28 | 177.5 ± 3.1   | 1.55 ± 0.56 |
| BL286AO | 2022-09-29 | 15 | E  | 770.6 ± 68.7 | 0 ± 0.01    | 0 ± 34        | 0.33 ± 0.02 |
|         |            |    | W  | 55.6 ± 26.9  | 0.62 ± 0.02 | −50.5 ± 1.7   | 0.27 ± 0.03 |
|         |            |    | S5 | 107.1 ± 25.6 | 0.95 ± 0.03 | −133.9 ± 2    | 0.38 ± 0.06 |
|         |            |    | S4 | 53.5 ± 18.8  | 1.47 ± 0.04 | −157 ± 1.5    | 0.43 ± 0.08 |
|         |            |    | S3 | 32.8 ± 15    | 2.04 ± 0.08 | −169.7 ± 2.2  | 0.61 ± 0.16 |
|         |            |    | S2 | 24.5 ± 14.9  | 4.22 ± 0.29 | 168.8 ± 4     | 1.08 ± 0.59 |
|         |            |    | S1 | 23.8 ± 16.3  | 5.67 ± 0.34 | 177.9 ± 3.5   | 1.38 ± 0.69 |
| BL286AU | 2023-05-27 | 15 | E  | 583.2 ± 53.3 | 0 ± 0.02    | 0 ± 20.6      | 0.41 ± 0.03 |
|         |            |    | W  | 32.1 ± 22.7  | 0.7 ± 0.03  | −44.8 ± 1.8   | 0.31 ± 0.03 |
|         |            |    | S5 | 111.9 ± 23.3 | 1.11 ± 0.03 | −139.1 ± 1.8  | 0.36 ± 0.06 |
|         |            |    | S4 | 23.8 ± 11.2  | 1.94 ± 0.08 | −168.4 ± 2.4  | 0.59 ± 0.16 |
|         |            |    | S3 | 9.4 ± 7.5    | 2.68 ± 0.18 | 178.4 ± 3.9   | 0.85 ± 0.36 |
|         |            |    | S2 | 22.9 ± 13.1  | 4.46 ± 0.35 | 168.8 ± 4.5   | 1.34 ± 0.7  |
|         |            |    | S1 | 17.7 ± 12.8  | 5.98 ± 0.51 | 178.7 ± 4.9   | 1.51 ± 1.02 |
| BL286BB | 2023-12-15 | 15 | E  | 662.8 ± 53.5 | 0 ± 0.01    | 0 ± 12.2      | 0.44 ± 0.03 |
|         |            |    | W  | 37.1 ± 24.3  | 0.64 ± 0.02 | −34.7 ± 1.6   | 0.28 ± 0.02 |
|         |            |    | S5 | 130.7 ± 23.5 | 1.14 ± 0.03 | −140.8 ± 1.4  | 0.38 ± 0.05 |
|         |            |    | S4 | 19.3 ± 9.4   | 1.97 ± 0.08 | −169.8 ± 2.4  | 0.56 ± 0.17 |
|         |            |    | S3 | 10.4 ± 7.2   | 2.82 ± 0.22 | −179.2 ± 4.4  | 0.84 ± 0.44 |
|         |            |    | S2 | 25.4 ± 13.9  | 4.42 ± 0.35 | 167.1 ± 4.5   | 1.36 ± 0.7  |
|         |            |    | S1 | 17.6 ± 12    | 6.02 ± 0.47 | 176.4 ± 4.4   | 1.46 ± 0.94 |
| BL286BG | 2024-06-02 | 15 | E  | 606.6 ± 53.3 | 0 ± 0.02    | 0 ± 18.6      | 0.48 ± 0.03 |
|         |            |    | W  | 36.7 ± 22.6  | 0.55 ± 0.02 | −41.2 ± 1.9   | 0.14 ± 0.01 |
|         |            |    | S5 | 126.3 ± 24.1 | 1.2 ± 0.03  | −146.4 ± 1.9  | 0.46 ± 0.07 |
|         |            |    | S4 | 9.5 ± 7.2    | 2.28 ± 0.12 | −171.6 ± 3    | 0.54 ± 0.23 |
|         |            |    | S3 | 14.4 ± 9.7   | 3.23 ± 0.36 | 174.2 ± 6.4   | 1.24 ± 0.73 |
|         |            |    | S2 | 18 ± 11.9    | 4.75 ± 0.38 | 165.9 ± 4.6   | 1.12 ± 0.76 |
|         |            |    | S1 | 14.4 ± 11.5  | 6.25 ± 0.63 | 177.7 ± 5.7   | 1.68 ± 1.26 |

**Note.** The columns are as follows: (1) project code, (2) observed epoch, (3) observed frequency, (4) component label, (5) average flux density, (6) radial distance from the E component, (7) position angle relative to the E component, and (8) size of the components.

**Table A2**
22 GHz Model Fitting Results

| P.C. | Date | Freq. (GHz) | ID | Flux Density (mJy) | Positions with Respect to E | | d (mas) |
| | | | | | r (mas) | PA (deg) | |
| (1) | (2) | (3) | (4) | (5) | (6) | (7) | (8) |
|---|---|---|---|---|---|---|---|
| BR145YB | 2013-11-30 | 22 | E | 493.5 ± 79.6 | 0 ± 0.01 | 0 ± 10.1 | 0.18 ± 0.02 |
| | | | W | 327.2 ± 65 | 0.29 ± 0.02 | −42.6 ± 4.1 | 0.25 ± 0.04 |
| | | | S2 | 86 ± 32.2 | 0.89 ± 0.05 | 174.7 ± 3.7 | 0.32 ± 0.1 |
| | | | S1 | 200.1 ± 94.1 | 1.98 ± 0.26 | −177.2 ± 7.7 | 1 ± 0.46 |
| BM421 | 2015-07-07 | 22 | E | 662 ± 60 | 0 ± 0.01 | 0 ± 5.5 | 0.15 ± 0.01 |
| | | | W | 240.2 ± 36.7 | 0.38 ± 0.02 | −25.9 ± 2.9 | 0.40 ± 0.04 |
| | | | S2 | 76.2 ± 20.8 | 1.03 ± 0.06 | 177.8 ± 3.4 | 0.51 ± 0.12 |
| | | | S1 | 71.4 ± 38.5 | 2.69 ± 0.46 | −177.8 ± 9.7 | 1.73 ± 0.92 |
| BR210CD | 2016-02-25 | 24 | E | 749.9 ± 67.4 | 0 ± 0.01 | 0 ± 15.5 | 0.21 ± 0.01 |
| | | | W | 69.7 ± 28.8 | 0.4 ± 0.01 | −59 ± 1.5 | 0.15 ± 0.02 |
| | | | S3 | 162.8 ± 33.6 | 0.58 ± 0.01 | 179.7 ± 0.8 | 0.09 ± 0.01 |
| | | | S2 | 44.5 ± 18.5 | 1.34 ± 0.1 | −178 ± 4.4 | 0.58 ± 0.2 |
| | | | S1 | 137 ± 96.6 | 3.07 ± 0.66 | −177.2 ± 12.1 | 1.88 ± 1.32 |
| UD001C | 2017-02-23 | 24 | E | 409.8 ± 58.5 | 0 ± 0.01 | 0 ± 42.9 | 0.2 ± 0.02 |
| | | | W | 77.6 ± 26.4 | 0.41 ± 0.01 | −56 ± 1.6 | 0.02 ± 0.01 |
| | | | S3 | 101.4 ± 36.1 | 0.67 ± 0.03 | −166.4 ± 3.1 | 0.25 ± 0.07 |
| | | | S2 | 116.3 ± 59.1 | 1.43 ± 0.19 | −179.1 ± 7.6 | 0.78 ± 0.38 |
| | | | S1 | 81.5 ± 79.1 | 3.61 ± 0.89 | −178.5 ± 13.9 | 1.86 ± 1.79 |
| UD001W | 2018-06-30 | 24 | E | 414.4 ± 61.7 | 0 ± 0.01 | 0 ± 29.7 | 0.19 ± 0.02 |
| | | | W | 76.4 ± 28.5 | 0.41 ± 0.02 | −64.7 ± 3 | 0.2 ± 0.04 |
| | | | S3 | 157.3 ± 43 | 0.74 ± 0.05 | −161.2 ± 3.8 | 0.42 ± 0.1 |
| | | | S2 | 24.5 ± 17.1 | 1.94 ± 0.14 | −178.6 ± 1.8 | 0.24 ± 0.12 |
| | | | S1 | 85 ± 83.9 | 3.24 ± 1.17 | −176.6 ± 19.9 | 2.39 ± 2.45 |
| BZ076A | 2019-03-17 | 22 | E | 335.6 ± 49.1 | 0 ± 0.01 | 0 ± 18 | 0.13 ± 0.01 |
| | | | W | 132.6 ± 31.1 | 0.43 ± 0.01 | −61.8 ± 1.7 | 0.13 ± 0.02 |
| | | | S3 | 40.8 ± 18.1 | 1.07 ± 0.07 | −152.5 ± 3.8 | 0.41 ± 0.14 |
| | | | S2 | 44.6 ± 23.1 | 2.08 ± 0.14 | −174.3 ± 4 | 0.61 ± 0.29 |

**Note.** The columns are as follows: (1) project code, (2) observed epoch, (3) observed frequency, (4) component label, (5) average flux density, (6) radial distance from the E component, (7) position angle relative to the E component, and (8) size of the components.

**Table A3**
43 GHz Model Fit Result

| P.C. | Date | Freq. (GHz) | ID | Flux Density (mJy) | Positions with Respect to E | | d (mas) |
| | | | | | r (mas) | PA (deg) | |
| (1) | (2) | (3) | (4) | (5) | (6) | (7) | (8) |
|---|---|---|---|---|---|---|---|
| BA111M | 2016-01-18 | 43 | E | 505 ± 49 | 0 ± 0.01 | 0 ± 21.1 | 0.10 ± 0.01 |
| | | | W | 89.4 ± 20.8 | 0.41 ± 0.01 | −55.7 ± 0.9 | 0.07 ± 0.01 |
| | | | S3 | 194.2 ± 28 | 0.42 ± 0.01 | −168.7 ± 1.2 | 0.16 ± 0.02 |
| | | | S2 | 33.7 ± 13.4 | 1.33 ± 0.15 | −178.1 ± 1.8 | 0.24 ± 0.08 |
| | | | S1 | 82.3 ± 85.6 | 2.72 ± 0.92 | −179.7 ± 18.7 | 1.78 ± 1.85 |
| BZ081A | 2021-04-22 | 43 | E | 312.8 ± 25 | 0 ± 0.01 | 0 ± 63.5 | 0.09 ± 0.01 |
| | | | W | 79 ± 12.7 | 0.51 ± 0.01 | −62.5 ± 0.7 | 0.10 ± 0.01 |
| | | | S5 | 22.7 ± 6.7 | 0.36 ± 0.02 | −116.3 ± 3.9 | 0.21 ± 0.05 |
| | | | S4 | 37 ± 9.3 | 0.79 ± 0.02 | −132.3 ± 2 | 0.25 ± 0.06 |
| | | | S3 | 40.9 ± 11.8 | 1.46 ± 0.05 | −153.5 ± 2.3 | 0.43 ± 0.12 |

**Note.** The columns are as follows: (1) project code, (2) observed epoch, (3) observed frequency, (4) component label, (5) average flux density, (6) radial distance from the E component, (7) position angle relative to the E component, and (8) size of the components.

## ORCID iDs

Haitian Shang https://orcid.org/0009-0002-8909-2935
Wei Zhao https://orcid.org/0000-0003-4478-2887
Xiaoyu Hong https://orcid.org/0000-0002-1992-5260
Leonid I. Gurvits https://orcid.org/0000-0002-0694-2459
Ailing Zeng https://orcid.org/0009-0000-9427-4608
Tao An https://orcid.org/0000-0003-4341-0029
Xiaopeng Cheng https://orcid.org/0000-0003-4407-9868